\documentclass[a4paper,twoside]{article}

\usepackage{epsfig}
\usepackage{subcaption}
\usepackage{calc}
\usepackage{amssymb}
\usepackage{amstext}
\usepackage{amsmath}
\usepackage{amsthm}
\usepackage{multicol}
\usepackage{pslatex}
\usepackage{apalike}
\usepackage{tikz}
\usepackage{algorithm2e}
\usepackage[bottom]{footmisc}
\usepackage{SCITEPRESS}     

\begin{document}

\title{Petri Net modeling of root hair response to phosphate starvation in \textit{Arabidopsis thaliana}}

\author{\authorname{Amber H.B. Fijn*\sup{1}, Casper H. Stiekema*\sup{1}, Stijn Boere*\sup{1}, Marijan Višić*\sup{1} and Lu Cao\sup{1}}
\affiliation{\sup{1}Leiden Institute of Advanced Computer Science, Leiden University, Einsteinweg 55, Leiden, The Netherlands}
\email{\{a.h.b.fijn, c.h.stiekema, s.boere.2, m.visic\}@umail.leidenuniv.nl, l.cao@liacs.leidenuniv.nl}
\textit{*These authors contributed equally to this study.}}


\abstract{Limited availability of inorganic phosphate (Pi) in soil is an important constraint to plant growth. In order to understand better the underlying mechanism of plant response to Pi, the response to phosphate starvation in \textit{Arabidopsis thaliana} was investigated through use of Petri Nets, a formal language suitable for bio-modeling. \textit{A. thaliana} displays a range of responses to deal with Pi starvation, but special attention was paid to root hair elongation in this study. A central player in the root hair pathway is the transcription factor ROOT HAIR DEFECTIVE 6-LIKE 4 (RSL4), which has been found to be upregulated during the Pi stress. A Petri Net was created which could simulate the gene regulatory networks responsible for the increase in root hair length, as well as the resulting increase in root hair length. Notably, discrepancies between the model and the literature suggested an important role for RSL2 in regulating RSL4. In the future, the net designed in the current study could be used as a platform to develop hypotheses about the interaction between RSL2 and RSL4.}

\onecolumn \maketitle \normalsize \setcounter{footnote}{0} \vfill

\section{\uppercase{Introduction}}
Inorganic phosphate (Pi) is essential for plant growth as plants incorporate it into their DNA and phospholipids. In addition, plants require it for biological processes such as photosynthesis \cite{Crombez,Kayoumu}. Limited (bio)availability of Pi in soil is an important constraint to plant growth and reduces crop yields worldwide \cite{Datta}. Common solutions to deal with this involve fertilization. However, fertilization is associated with sustainability issues as it can deteriorate soil ecology \cite{Hopkins}. In this context, researchers have focused efforts on understanding and elucidating the molecular mechanisms underlying plant response to Pi starvation. By doing this, researchers hope to identify traits that can be manipulated to develop cultivars with enhanced resilience to Pi starvation. It has been well described that plants display a suit of responses to deal with limited Pi availability. For example, to increase the bioavailability of Pi in soil, plants secrete acid phosphatases and ribonucleases which release P esters bound to soil components \cite{Abel,Bariola}. \\
However, the most distinct response can be found in the roots. Plants display both an increased lateral root formation as well as increased root hair elongation under Pi stress \cite{Crombez}. Increase in root hair length and density has been reported to account for 90\% of the total Pi uptake during Pi stress \cite{Fohse}. In this case study, we aimed to gain further insight into the root hair response to phosphate (Pi) starvation in \textit{A. thaliana}.  As most research focuses on single genes, we hoped that by incorporating these results into a single Petri Net further insight could be gained into the importance and role of the regulatory network components. We focused our research on \textit{A. thaliana} as the Pi response has been well-studied in this plant model. \\
Petri Nets can be applied to model concurrent, asynchronous, and stochastic systems \cite{Murata}. Hence, Petri Nets have been used as a graphical and mathematical tool for quantitative and/or qualitative analyses of complex biological systems. Petri Nets represent a directed, weighted, bipartite graph with two types of nodes: places and transitions which are connected by arcs. This framework has been successfully applied to various biological processes including biochemical pathways, signal transduction as well as epidemic and ecological models \cite{Chaouiya,Hardy,terberov2024}. 

\section{\uppercase{Biological Background}}
\begin{figure}[h!]
\centering
\includegraphics[width=\linewidth]{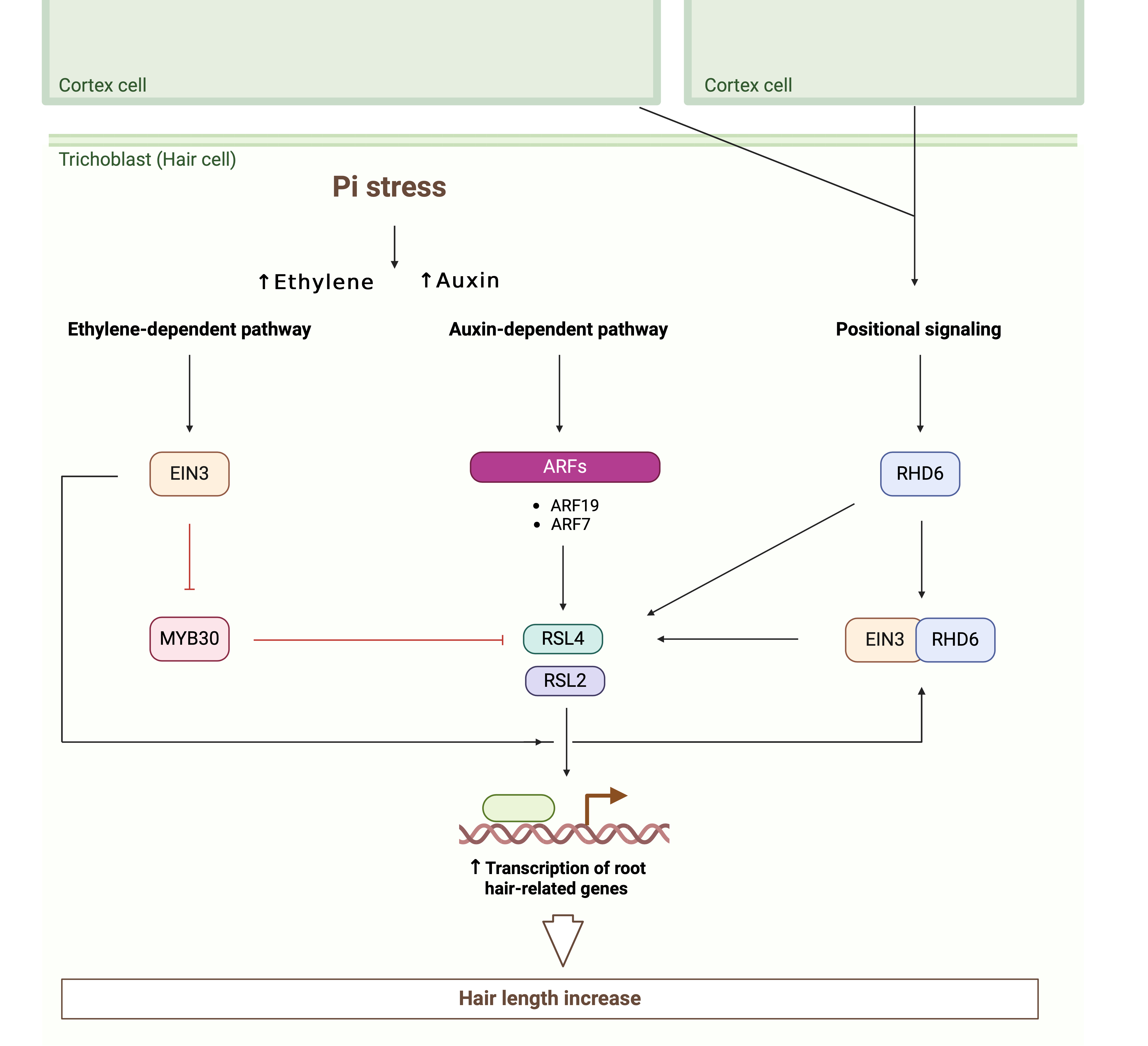}
    \caption{Overview of regulatory networks towards root hair elongation. Figure created in BioRender.}
    \label{fig:BioR}
\end{figure}

Many genes are involved in plant response to root hair elongation. Among these, the transcriptional regulator ROOT HAIR DEFECTIVE 6-LIKE 4 (RSL4) is recognized as a central player in the network as it activates many downstream targets essential for root morphogenesis \cite{Bhosale}. In addition, it has been found that RSL4 levels increase during Pi stress, and knockout of RSL4 significantly impairs root hair elongation. Besides RSL4, other transcription factors such as ROOT HAIR DEFECTIVE 6 (RHD6) and ROOT HAIR DEFECTIVE 6-LIKE 2 (RSL2) have also been highlighted as key genes involved in root hair development. Regulation of these transcription factors occurs in complex networks that also involve plant hormones. Auxin has been found to modulate RSL4 levels \cite{Yi}. Auxin achieves this by activating auxin response factors (ARFs) that regulate the expression of RSL2 and RSL4 \cite{Bhosale}. Besides auxin, ethylene is also involved in the root hair response to Pi stress \cite{Song}. Intracellular reception of ethylene increases the levels of transcription factor ETHYLENE INSENSITIVE 3 (EIN3). Similar to the ARFs, EIN3 can regulate RSL4 expression and, in addition, has been found to share gene targets with RSL4.\\
In the root epidermis, root hair development of epidermal cells is tightly regulated \cite{Montiel}. Epidermal cells receiving a positional signal from two cortical cells develop a root hair (trichoblasts), whereas cells receiving a positional signal from a single cortical cell remain hairless (atrichoblasts). To commit to the trichoblast fate a process of lateral inhibition is essential \cite{Savage}. Differentiating atrichoblast cells express CAPRICE (CPC). The CPC proteins, among other proteins, move to adjacent trichoblast cells where they form a complex with GLABRA3 (GL3) and TRANSPARENT TESTA GLABRA1 (TTG1). This complex allows the differentiating trichoblast cell to remain in the default, trichoblast pathway. In addition, CPC is essential for the expression of the transcription factor RHD6 which is involved in the early stages of root hair development \cite{Salazar}. RSL4 is one of its targets and is necessary for the transcription of genes required for root hair formation. RHD6 was also found to control RSL2, which like RSL4 controls downstream targets involved in root hair morphogenesis, indicating some redundancy between the two \cite{Mangano}. Under Pi starvation, the plant hormones auxin and ethylene play a vital role in enhanced root hair elongation \cite{Vissenberg}. The networks involved in this hormone-dependent response to Pi stress will be discussed in the following sections and an overview can be found in Figure \ref{fig:BioR}. Together, these networks result in a 1.7-fold increase in root hair length from Pi sufficient to Pi deficient conditions \cite{Datta2}. 
\subsection{Auxin-dependent pathway}
It has been reported that auxin levels increase 2.5-fold in response to Pi stress \cite{Bhosale}. Increased auxin biosynthesis occurs in the root tip and is dependent on the gene TRYPTOPHAN AMINOTRANSFERASE OF ARABIDOPSIS1 (TAA1), which is upregulated upon Pi stress. The auxin transporter AUXIN RESISTANT 1 (AUX1) is responsible for the subsequent transport of auxin towards the root hair elongation zone. Once auxin reaches the root hair zone, it causes enhanced activity of the auxin-inducible transcription factors ARF7 and ARF19. For ARF19 a 2-fold increase of transcript levels has been described \cite{Bhosale}. ARF19 can activate the expression of RSL4 and RSL2, whereas ARF7 can activate only RSL4. Together with other transcription factors, this leads to 1.8-fold and 2.3-fold changes of RSL4 and RSL2 mRNA, respectively, during Pi starvation.

\subsection{Ethylene-dependent pathway}
Changes in ethylene levels during Pi starvation are less well described and no clear effect has been elucidated \cite{Roldan}. Nonetheless, it has been found that, during Pi stress, levels of the ethylene-response factor EIN3 are increased 2-fold \cite{Song}. Various functions in regulating root hair response in low Pi have been ascribed to EIN3. Among these, transcriptional regulation of various genes is involved in root hair development. Many of these genes are also regulated by RSL4, highlighting the redundancy present in the gene regulatory network. In addition, EIN3 is involved in the regulation of RSL4 itself, through inhibition of MYB30 \cite{Xiao}. MYB30 can bind to the promoter region of RSL4 and inhibit its expression. Ethylene enhances the physical association of EIN3 with MYB30 to form the EIN3-MYB30 complex. This effectively reduces the association of MYB30 with the RSL4 promoter and thus allows for increased expression of RSL4. Finally, EIN3 also physically interacts with RHD6, and the transcription factors cooperatively bind to the promoter region of RSL4 with increased affinity compared to the binding affinity when RHD6 binds by itself \cite{Feng}.

\subsection{RSL4 regulation}
In addition to the hormone-mediated regulation of RSL4 expression, several independent regulatory loops have been found crucial for the fine-tuning of timing and amount of total RSL4 product. RSL4 has been shown to bind to its own promoter and enhance transcription, forming a positive feedback loop \cite{Hwang}. GT-2-LIKE-1 (GTL1) and its less studied homolog DF1 form a negative feedback loop with RSL4 on a transcriptional level \cite{Shibata}. Expression of GTL1 is initiated by the RSL4 protein, after which GTL1 binds to the promoter of RSL4, causing repression of transcription. GTL1 also represses gene expression downstream from RSL4 and its expression. An additional negative feedback loop, centered around LONG ROOT HAIR (LRH), is present on the translational level \cite{Cui}. RSL4 initiates the transcription of LRH, which then binds to the eukaryotic TRANSLATION INITIATION FACTOR 4E (eIF4E). eIF4E recognizes the 5’ cap of the mature mRNA molecules and helps with the ribosome binding. However, by binding to LRH, eIF4Es lose the affinity to associate with the mRNA of RSL4, effectively inhibiting translation. Besides the pathways mentioned here, RSL4 is regulated by many other genes and complexes, including ZINC-FINGER PROTEIN 1 (ZP1) \cite{Han} and RALF1-FERONIA complex \cite{Zhu}, but the details about their impact on total expression are less clear. The result of this complex regulation is a sequence of events determining the final length of the hair \cite{Datta2}. RSL4 protein starts being synthesized 2 hours before the root hair growth initiation and achieves its highest abundance 2 hours after the initiation. The expression of RSL4 mRNA is detectable only before initiation of hair growth. The abundance of RSL4 protein starts to decrease gradually over several hours during the elongation phase due to degradation mediated by proteasome 26S \cite{Datta2}. As long as the RSL4 protein is present in the trichoblasts, the hair keeps growing at a steady rate of 1 micrometer per minute \cite{Datta2,Yi}.

\section{\uppercase{Petri Nets modeling}}
\begin{figure}[h!]   
    \begin{center}
    \includegraphics[width=0.3\textwidth]{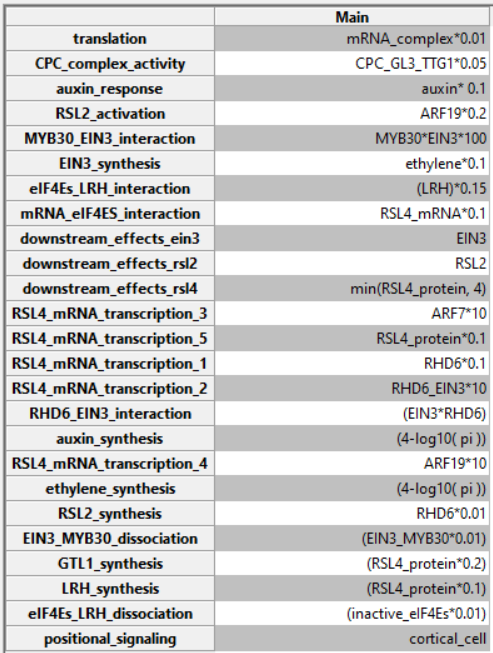}
    \caption{The functions in Petri Net transitions.}
    \label{table:functions}
    \end{center}
\end{figure}

\begin{figure}[h!]
\centering
\begin{tikzpicture}
    \node[anchor=south west, inner sep=0] (image) at (0,0) {\includegraphics[trim={0.5cm 0 0.5cm 0}, clip, width=\linewidth]{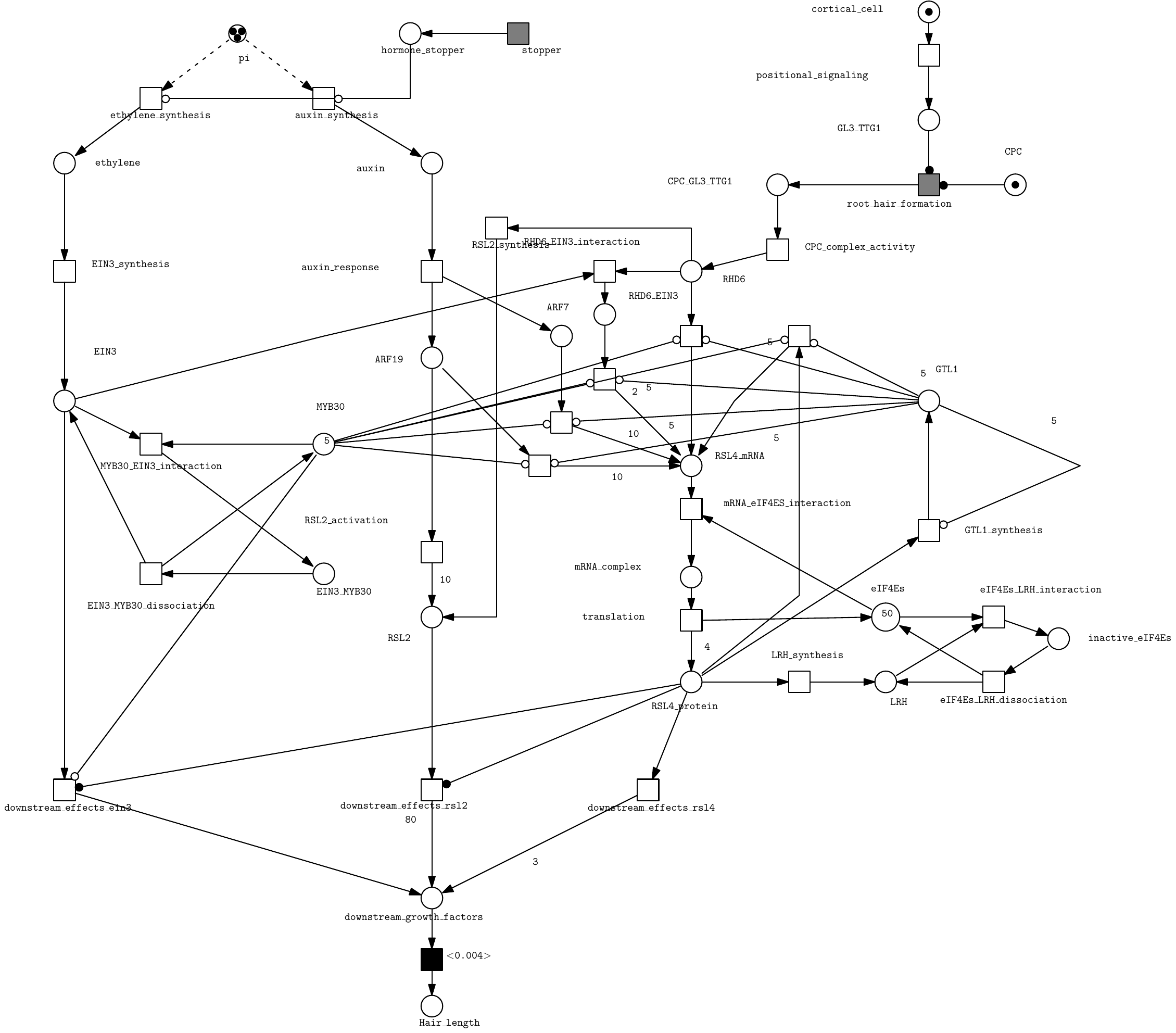}};
    \begin{scope}[x={(image.south east)},y={(image.south west)}]
        \node[anchor=south west] at (0.7, 0.7) {\includegraphics[width=0.3\linewidth]{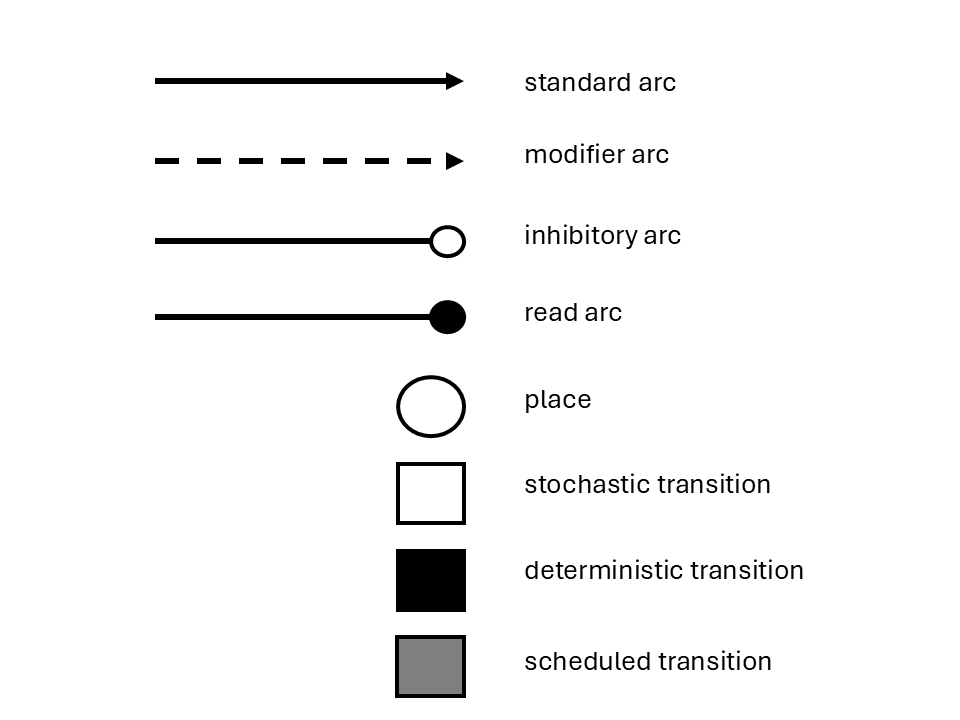}};
    \end{scope}
\end{tikzpicture}
\caption{The Petri Net model of root hair elongation in response to inorganic phosphate (Pi).}
\label{fig:wfc}
\end{figure}

Petri Nets are modeling structures based on mathematical formalisms and are suitable for describing complex systems with multiple co-occurring and co-dependent processes. A Qualitative Petri Net (QPN) is defined by the following quadruple $N = (P, T, f, m_{0})$ \cite{Heiner}. Here, P refers to a set of places (represented by circles), and T to a set of transitions (represented by rectangles). $f$ defines the set of directed arcs that connect transitions and places, which can be weighted by non-negative integers, and finally $m_{0}$ which gives the initial marking. Places are typically used to model passive system components, which in a biological context can be used to represent species, proteins, or molecules. The available amount of these compounds or species is represented by tokens, shown as black dots or non-negative integers. Transitions are active components representing processes such as chemical reactions or interactions. Arcs connect places and transitions \cite{Heiner,Blatke}. The tokens assigned to the places before the first firing is the initial marking. Tokens move through the net by firing transitions \cite{Blatke}. When a transition is fired, the tokens are removed from the pre-place(s) and added to the post-place(s). The number of moving tokens is determined by the weights of the ingoing and outgoing arcs of a transition. Firing happens atomically and consumes no time in the QPN \cite{Heiner}.\\ 
In this model, we used a Stochastic Petri Net, where a firing delay of a transition is randomly computed based on the probability distribution with a firing rate. This allows the incorporation of time property and randomness effect. They are valuable for modeling biological processes. Most functions in our model are based on mass action kinetics, where the firing rate of a transition depends on the number of tokens in its pre-place(s), multiplied by a constant that was determined empirically for every transition. The use of alternative functions, namely logarithmic and minimum functions, is motivated in subsequent sections. Besides stochastic transitions, there are deterministic transitions that fire without delay and are only active during certain simulation periods.\\ 
Tokens in the net represent both information and actual amount. For example, the number of tokens of Pi represents the concentration (µM) of Pi in sufficient and deficient conditions. For transcription factor activation, tokens represent information, as information from the transcription factor is transferred to its activated gene. In order to allow the Pi signal to directly affect the abundance of proteins and the eventual root hair length, most places had zero initial marking. Only MYB30 and eIF4E were given non-zero initial marking since both components are already present in the cell. The model can be interpreted as the intracellular processes in response to a specific Pi concentration. Two Pi concentrations will be investigated, 3 and 300 µM. These were concentrations used to represent Pi deficient and sufficient conditions.\\
In a Petri Net several additional types of arcs can be defined \cite{Marwan,Blatke}. With read arcs, the transition fires only when its pre-place is populated with an equal or larger number of tokens as its associated arc weight. Tokens are not consumed from the pre-place. Inhibitor arcs prevent a transition as long as its pre-place holds an equal or larger number of tokens to the arc weight. Modifier arcs are used with stochastic transitions and attribute to a token-dependent modification of the firing rate. No tokens are consumed from the pre-place.\\
To create the Petri Net, the Snoopy software was employed \cite{Snoopy}. The complete Petri Net can be seen in Figure \ref{fig:wfc}, with the initial marking representing low Pi configurations. In the model, custom functions were ascribed to each transition and designed to be in line with the biological system. In Figure \ref{table:functions} the functions used for each transition are displayed.
\\

The epidermal cell resolution based on positional signaling was modeled by including the components necessary for determining trichoblast cell fate. Positional signaling from cortical cells and CPC originating from adjacent atrichoblasts were required for the production of RHD6 and for the cell to commit to trichoblast cell fate.

\subsection{Modeling auxin-dependent pathway}
Pi influences auxin levels through a modifier arc with the function $(4-log10(pi))$ which causes higher levels of Pi to be associated with a lower firing rate. The log function was selected as a log relation has been described between Pi concentration and root hair length \cite{Bates}. Since experimental Pi levels in literature do not go above 1000 µM, the upper log-scaled limit was set to one order of magnitude more, which is 4. The modifier arc ensures that tokens are not consumed, resulting in the hair elongation being limited only by the marking-dependent firing rate. This was a simplified method to represent the Pi concentration which is unlikely to change much during root hair elongation. Auxin response was modeled through ARF7 and ARF19, which affect RSL2 and RSL4. The firing rate functions for these transitions differ, with the coefficients determined by empirical evaluation to keep the ratios between the downstream changes caused by these proteins in alignment with the literature.

\subsection{Modeling ethylene-dependent pathway}
The ethylene-dependent pathway begins with a modifier arc originating from Pi and includes a transition that affects ethylene levels in the same manner as it does for auxin. The effect of ethylene is mediated by EIN3. EIN3 binds to RHD6 and increases RSL4 transcription. As can be seen from its function, EIN3-RHD6 increases the RSL4 mRNA level more strongly than RHD6 alone. In addition, EIN3 can bind to MYB30. As this transition consumes tokens from MYB30, it can disrupt the inhibition of MYB30 on RSL4 transcription. As proteins are dynamic and are not infinite in a complex once bound, a loop was created to display the dissociation of the EIN3-MYB30 complex.

\subsection{Modeling RSL4 regulation}
RSL4 is produced in a pulse during root hair elongation. Two feedback loops responsible for the pulse are included in the net. The first feedback loop is the GTL1 loop. Production of the RSL4 protein leads to GTL1 production which inhibits transcription of RSL4 mRNA. This inhibitory loop only occurs when sufficient RSL4 has been produced. Moreover, since GTL1 also inhibits itself, a self-inhibitory arc was added. The second is LRH-mediated negative feedback loop. In the net, eIF4Es have a marking of 50, meaning that only when sufficient RSL4 has been produced, translation is fully inhibited. Dissociation of eIF4E and LRH can make eIF4E available again, reproducing the dynamic process of protein interactions in the cell. A simplification was made by marking downstream growth factors as a single place because the population of downstream growth factors was too large to model and limited information is known. Since RSL4 has been highlighted as a necessary transcription factor for root hair elongation, read arcs are included from RSL4 to transitions coming from EIN3 and RSL2. This ensures that, without RSL4, root hair elongation is not possible. To decrease the effect of the abundance of RSL4 on hair growth rate, the coefficient for firing rate was limited to 4 using the minimum function. From RSL4, RSL2, and EIN3, downstream growth factors are activated. These are simplified as a single place. From downstream growth factors, a deterministic transition is modeled towards the final hair length so that the elongation period becomes dependent on the number of tokens in pre-place with the rate held constant.

\section{\uppercase{Results}}
\subsection{Snoopy Simulation}
Using Snoopy simulation, a comparison was made for the root hair length, as well as protein and hormone abundances under low and high Pi concentrations. All graphs obtained from simulations are an averaged result of 100 separate runs.

\begin{figure}[h!]
    \centering
    \begin{subfigure}[b]{0.4\linewidth}
        \includegraphics[width=\linewidth]{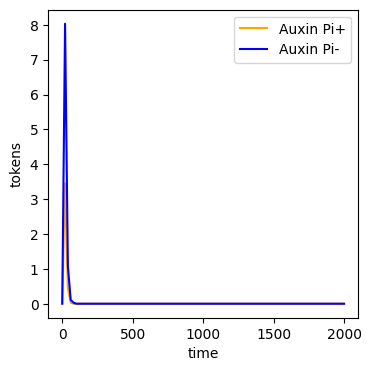}
        \caption{Auxin}
        \label{fig:auxin}
    \end{subfigure}
    \begin{subfigure}[b]{0.4\linewidth}
        \includegraphics[width=\linewidth]{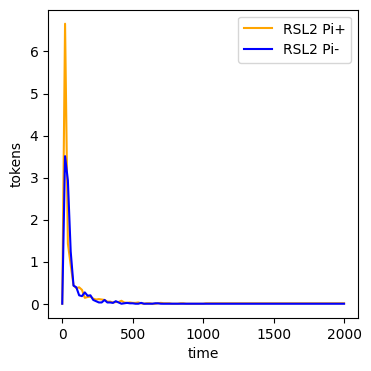}
        \caption{RSL2}
        \label{fig:RSL2}
    \end{subfigure}
    \begin{subfigure}[b]{0.4\linewidth}
        \includegraphics[width=\linewidth]{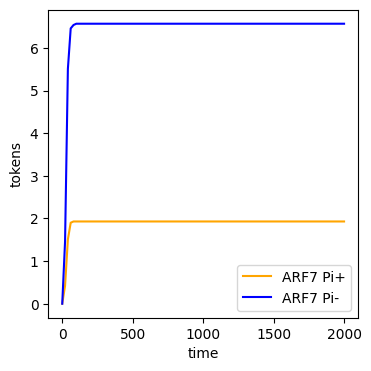}
        \caption{ARF7}
        \label{fig:ARF7}
    \end{subfigure}
    \begin{subfigure}[b]{0.4\linewidth}
        \includegraphics[width=\linewidth]{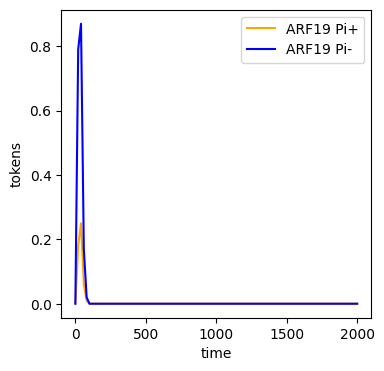}
        \caption{ARF19}
        \label{fig:ARF19}
    \end{subfigure}
    \begin{subfigure}[b]{0.4\linewidth}
        \includegraphics[width=\linewidth]{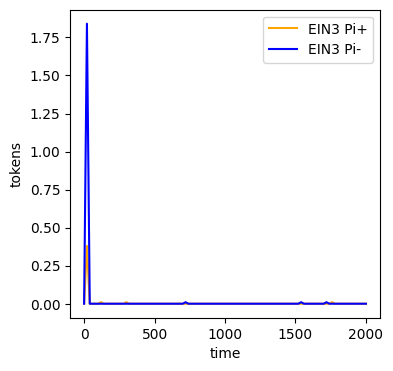}
        \caption{EIN3}
        \label{fig:EIN3}
    \end{subfigure}
    \begin{subfigure}[b]{0.4\linewidth}
        \includegraphics[width=\linewidth]{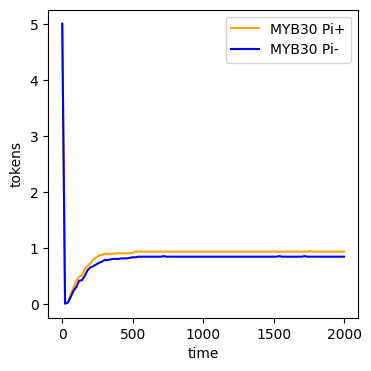}
        \caption{MYB30}
        \label{fig:MYB30}
    \end{subfigure}
    \begin{subfigure}[b]{0.4\linewidth}
        \includegraphics[width=\linewidth]{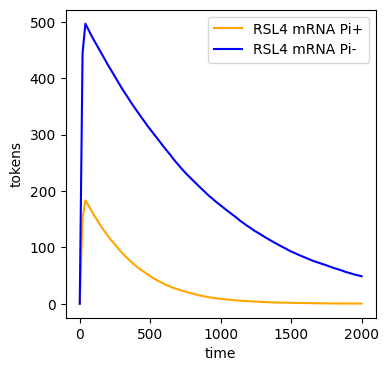}
        \caption{RSL4 mRNA}
        \label{fig:RSL4mRNA}
    \end{subfigure}
    \begin{subfigure}[b]{0.4\linewidth}
        \includegraphics[width=\linewidth]{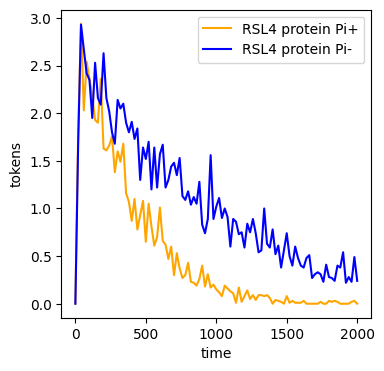}
        \caption{RSL4 protein}
        \label{fig:RSL4protein}
    \end{subfigure}

\caption{Auxin (a), RSL2 (b), ARF7 (c), ARF19 (d), EIN3 (e), MYB30 (f), RSL4 mRNA (g) and RSL4 protein (h) given a low (Pi-) and a high (Pi+) concentration of Pi.}
\end{figure}

A decrease in Pi concentration causes increased auxin biosynthesis. Figure \ref{fig:auxin} shows the presence of auxin in both low and high Pi concentrations. Auxin increases activity of ARF7 and ARF19 as shown in Figure \ref{fig:ARF7} and \ref{fig:ARF19}. The abundance of RSL2 in low and high Pi concentrations is shown in Figure \ref{fig:RSL2}. Under the influence of ARF19, RSL2 increases the transcription of downstream growth factors. However, this transcription also requires the presence of RSL4. Since the transcription of RSL2 in the auxin-dependent pathway occurs faster than the translation of RSL4, many tokens accumulate in RSL2 before RSL4 allowing them to flux towards the transcription of downstream growth factors. This illustrates why a peak can be observed for RSL2 abundance over time. The peak is lower in low Pi concentration due to higher RSL4 protein production, which leads to an earlier flux of RSL2 towards downstream growth factors.

The abundance of ethylene follows the same trend as auxin. Lower Pi leads to higher EIN3 levels as shown in Figure \ref{fig:EIN3}. EIN3 can physically interact with RSL4-inhibitor MYB30, and thus effectively release RSL4 from this inhibition. Figure \ref{fig:MYB30} shows the presence of MYB30 given both low and high Pi concentrations. The abundance of MYB30 quickly drops for both Pi concentrations, followed by a gradual increase and leveling out. This is due to the termination effect exerted on the ethylene upstream, which causes EIN3 to decline and allows MYB30 to become available again.
\\

As evident from the Figure \ref{fig:RSL4mRNA}, lower Pi concentration leads to a higher accumulation of RSL4. This is due to the culmination of activities from various proteins and hormones that are influenced by Pi. The general curve-shape of RSL4 abundance can be attributed to the inhibitory effect from GTL1, which inhibits transcription of RSL4 after some time, and subsequent RSL4 translation as shown in Figure \ref{fig:RSL4protein}. The fluctuating levels observed for RSL4 protein are caused by eIF4E, which is required for translation of RSL4 mRNA. After a large amount of RSL4 is translated, eIF4E is shortly disabled which causes a dip in the abundance of RSL4 protein, as the protein is still fluxed towards root hair elongation. Once eIF4E is enabled, an increase in RSL4 protein occurs, once again disabling eIF4E. Although the peak amount of tokens in the place representing RSL4 protein is similar in both Pi conditions, this should not be interpreted as RSL4 being produced in equal amounts throughout the simulation because it is constantly used by downstream processes. The slower leveling out in lower Pi conditions shows that RSL4 persists for a longer time in the cell, resulting in longer hairs.\\

\begin{figure}[!h]
    \centering
    \hspace*{-1.0cm}
    \includegraphics[width=0.7\linewidth]{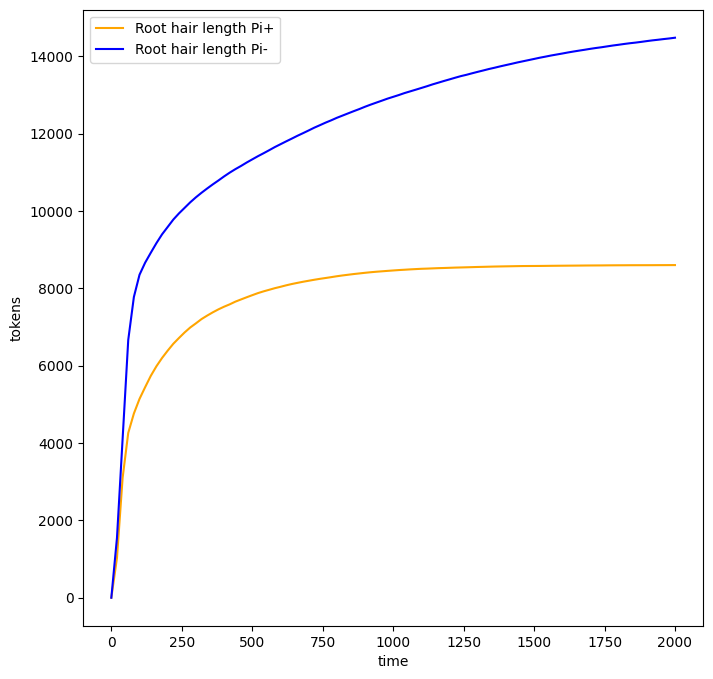}
    \caption{Root hair length given a low and a high concentration of Pi.}
    \label{fig:hairlength}
\end{figure}

The transcription of the downstream growth factors is influenced by three factors: RSL4 protein which is influenced by RSL4 regulation as well as the ethylene- and auxin-dependent pathway, RSL2 which is produced through the auxin-dependent pathway, and finally EIN3 which is influenced by the ethylene-dependent pathway. Figure \ref{fig:hairlength} shows the hair length given a low and a high concentration of Pi. While Pi cannot directly influence hair length, the previous figures indicated the effect that Pi has on the proteins and hormones, which eventually determine the root hair length. The hair growth does not follow a steady rate as described in the literature. Instead, a high initial growth rate can be observed, likely due to the high number of tokens accumulated in RSL2, and to a lesser extent EIN3, that cannot immediately flow towards downstream growth factors because RSL4 protein is required through the read arcs.

\subsection{Validation}
For various hormones and proteins, literature has reported fold changes in their abundance given a lower or higher concentration of Pi. To validate our network we compare the results between low and high Pi concentration simulations to those obtained from literature.
\begin{table}[h!]
    \centering
    \begin{tabular}{c|c|c}
	& Literature & Our Petri Net\\
\hline
RSL4 mRNA	        &	1.8 & 2.3  \\
RSL4 protein        &	2.5 & 2.0    \\
RSL2                &	2.3 & 1.5    \\
ARF19	            &	2.0 & 2.3    \\
Auxin               &	2.5 & 2.3    \\
EIN3                &	2.0 & 2.3    \\
Root hair length	&	1.7 & 1.7    \\
    \end{tabular}
    \caption{The relative abundance of hormones and proteins in low Pi conditions compared to a high Pi conditions.}
    \label{tab:compare}
\end{table}

Table \ref{tab:compare} shows that our model achieves comparable results to some protein and hormone values obtained from the literature (e.g., ARF19, EIN3) yet lacks in accuracy when compared to others. For example, the result obtained for RSL2 appears biologically out of bounds when considering the respective literature \cite{Bhosale}. Quantitative inaccuracies are due to incapable of capturing the intricate interplay between these proteins and hormones from limited available knowledge in general. Moreover, the validation set was not obtained from a single study, which could have affected the compatibility of simulating these in a single model. 

\section{\uppercase{Conclusion}}
We developed a Petri Net to simulate the regulatory network governing root hair length in \textit{A. thaliana}, with specific attention to soil Pi availability. Our model, based on stochastic properties of biological processes, incorporates key regulatory components of the system, resulting in outputs that closely align with experimental data reported in the literature. This includes the relative size of root hair, the amount of hormones and their response factors in the cell, and the pulse-like expression of the key regulator RSL4, including both its mRNA and protein. Given the scarcity of review articles on this topic, our model advances the current understanding of this biological process. It also highlights critical gaps in the existing knowledge. Further research should focus on the regulation of RSL2, as understanding RSL4 regulation alone is insufficient to fully replicate all aspects of root hair growth. Obtaining these data could aid in addressing some of the quantitative inaccuracies. Additionally, the permanent transcriptional repression of RSL4 should be further investigated, and it may be linked to epigenetic modifications initiated by downstream genes. Due to the lack of this information, we had to opt for a simplistic approach of using a scheduled transition to prevent infinite Pi response. This data could additionally provide the information needed for better timing of the net, as we were not able to fully replicate the sequence of the events in time. Additional data needed to circumvent these discrepancies includes the rate of proteasomal degradation of RSL4, relative binding affinities of RSL4 to different components, and the stability of RSL4 mRNA. Despite this, the key features of the process are evident. This is particularly true for the expression of RSL4 mRNA in the initial stage, while RSL4 protein is synthesized over longer period of time but with gradual decrease in abundance.\\
Our model represents an initial step towards constructing a comprehensive plant root system model. Future enhancements should incorporate missing genetic data and account for responses to other soil nutrients, such as iron and bioavailable nitrogen, given their interdependent effects \cite{Crombez,Liu}. Integrating this genetic model with morphological root models, such as SimRoot \cite{Postma}, could yield a robust exploratory tool for simulating plant responses to diverse ecological variables. Considering the conserved nature of Pi sensing mechanisms between the model plant \textit{A. thaliana} and other plants, including crops like rice and maize \cite{Hwang,Ren}, comprehensive models could become perfect tools for accelerating the development of plants with optimized agronomic traits.

\bibliographystyle{apalike}
{\small
\bibliography{Paper}}

\end{document}